\documentclass[aps,epsfig,twocolumn]{revtex4}
\usepackage{amsmath}
\usepackage{epsfig}
\usepackage{graphics}
\begin{document}
\title{Viscosity and Thermal Relaxation for a resonantly interacting Fermi gas}

\author{G.\ M.\ Bruun}

\affiliation{Niels Bohr Institute, Blegdamsvej 17, DK-2100
Copenhagen \O, Denmark}

\author{H.\ Smith}

\affiliation{\O rsted Laboratory, Niels Bohr Institute,
Universitetsparken 5, DK-2100 Copenhagen \O, Denmark.}

\date{\today{}}

\begin{abstract}
The viscous and thermal relaxation rates of an interacting fermion
gas are calculated as functions of temperature and scattering
length, using a many-body scattering matrix which incorporates
medium effects due to Fermi blocking of intermediate states. These
effects are demonstrated to be large close to the transition
temperature $T_c$ to the superfluid state. For a homogeneous gas
in the unitarity limit, the relaxation rates are increased by
nearly an order of magnitude compared to their value obtained in
the absence of medium effects due to the Cooper instability at
$T_c$. For trapped gases the corresponding ratio is found to be
about three due to the averaging over the inhomogeneous density
distribution. The effect of superfluidity below $T_c$ is
considered  to leading order in the ratio between the energy gap
and the transition temperature.
\end{abstract}

\maketitle

Pacs Numbers: 03.75.Ss, 05.30.Fk,51.20.+d

\section{Introduction}

The properties of resonantly interacting fermions have been under
intense investigation during the last few years, stimulated by the
ability to manipulate  the interatomic interaction through the use
of  Feshbach resonances. In this way it has been possible to
create molecular Bose-Einstein condensates from weakly bound
molecules consisting of two  fermionic atoms. The molecular
condensates are formed when the scattering length is positive,
corresponding to an effective repulsion between particles. If the
interaction is attractive (the scattering length negative) the
atoms may pair up in a manner similar to the way in which
electrons form Cooper pairs in superconducting metals.  The
important dimensionless parameter for an interacting, degenerate
fermion gas is the product of the Fermi wave number $k_{\rm F}$,
given by $k_{\rm F}^3=3\pi^2n$ where $n$ is the fermion density,
and the scattering length $a$. For attractive interactions, and
for $k_{\rm F}|a|$ much less than unity (which we refer to as the
weak-coupling limit), the transition temperature $T_c$ to the
superfluid phase is very much less than the Fermi temperature
$T_{\rm F}$. Near Feshbach resonances it is possible to achieve
values of $k_{\rm F}|a|$ that are much larger than unity (the
strong-coupling limit), and in this case the expected transition
temperature is comparable to the Fermi temperature, typically
about one quarter of $T_{\rm F}$, depending on the model used for
the effective scattering matrix.

In this work we demonstrate  how many-body effects have a
significant influence on the viscous and thermal relaxation rates
at temperatures near the transition temperature to the superfluid
state. The many-body effects are due to Fermi blocking of
intermediate states in the effective scattering matrix. Such
effects are known to influence equilibrium quantities such as
chemical potential and total energy, but - as we shall see below -
their influence on properties like thermal or viscous relaxation
is much more pronounced. Thermal relaxation rates may be
determined directly experimentally~\cite{Loftus,Regal}, and
viscous relaxation rates can be extracted from the attenuation of
collective modes. Our work is an extension of~\cite{Massignan}, in
which the viscous relaxation rate was calculated using the vacuum
scattering matrix, and we shall refer to this paper as I in the
following.

\section{Viscosity and Thermal Relaxation}\label{Viscsection}

In this section  we discuss the relation between the thermal and
viscous relaxation rates for a two-component Fermi gas trapped in
a harmonic-oscillator potential, with an equal number of particles
in each component. The interaction between the
fermions is taken  to be attractive, $a<0$, and we consider
fermions in their normal state, at $T>T_c$. For bosons in a
harmonic trap, at temperatures above the Bose-Einstein transition
temperature, the thermal and viscous relaxation rates have been
shown to be  simply related, differing only by an overall factor
of two~\cite{PethickSmithBook}. As one might expect, the same
relation applies to the fermion gas, since the factor of two
originates in the equality of kinetic and potential energies for a
particle in a harmonic trap. We briefly indicate the line of
reasoning which leads to this result and refer the reader
to~\cite{PethickSmithBook} for details.

We consider a gas of particles with mass $m$, trapped in an axially-symmetric
harmonic-oscillator potential of the form
\begin{equation}
V({\bf r})=\frac{m}{2}(\omega_{\perp}^2r_{\perp}^2+\omega_z^2z^2),
\end{equation}
where $r_{\perp}^2=x^2+y^2$. For such a system we define the
spatially averaged viscous relaxation rate by
\begin{equation}
 \frac{1}{\tau_\eta}=\frac{\int d^3r( X,H[X])}{\int d^3r(
 X,X)},
\label{avtaueta}
\end{equation}
where the symbols have the same meaning as in I. In particular $H$
is an integral operator derived from the collision integral of the
Boltzmann equation, and $X$ is a trial function given by
\begin{equation}
X=v_yp_x\label{X},
\end{equation}
with ${\bf p}$ being the particle momentum and ${\bf v}$ the group
velocity, equal to ${\bf p}/m$ in the normal state. In carrying
out the spatial average we treat the trapping potential in the
Thomas-Fermi approximation via a locally varying chemical
potential $\mu({\bf r})=\mu-V({\bf r})$. The time $\tau_\eta$
determines the nature and the damping of the collective modes in
the trapped atomic gases~\cite{Massignan}.

The measurement of  thermal relaxation rates has proven a useful
tool for investigating the  properties of trapped
gases~\cite{Loftus,Regal}. In such an experiment, the gas is
``heated'' preferentially in one spatial direction, and the
relaxation toward equilibrium with a uniform temperature in all
directions is determined from the time evolution of the rms cloud
radii.  The thermal relaxation rate can  be extracted from the
time-dependent  aspect ratio. In analogy with the viscous
relaxation rate, one may obtain a variational expression for the
relaxation time of temperature anisotropies in a
trap~\cite{PethickSmithBook},
\begin{equation}
\frac{1}{\tau_T}= \frac{\int
d^3r(\Phi_T,H[\Phi_T])}{\int
d^3r(\Phi_T,\Phi_T)}, \label{ThermalRate}
\end{equation}
where the trial  function is
\begin{equation}
\Phi_T=p_z^2-p^2/3+\frac{m^2}{3}(2\omega_z^2z^2-\omega_{\perp}^2r_{\perp}^2).\label{Phi}
\end{equation}
The momentum-dependent part of this trial function involves only a
$l=2$ spherical harmonic in momentum space. If the spatial part of
the trial function were neglected, the viscous and thermal
relaxation rates would be identical, since the trial function
(\ref{X})  also involves only $l=2$ spherical harmonics. Since
collisions relax only momentum anisotropies, the spatial part of
(\ref{Phi}) does not contribute to the numerator in
(\ref{ThermalRate}), whereas it gives the same contribution  to
the denominator as the momentum part. It therefore follows from
comparing (\ref{avtaueta}) and (\ref{ThermalRate}) that the
thermal  and  viscous relaxation rates are related by
\begin{equation}
\tau_T=2\tau_\eta.\label{TheSame}
\end{equation}
The viscous relaxation rate is thus directly accessible
experimentally via  measurements of the thermal relaxation rate as
in \cite{Loftus,Regal}.

\section{Collisions}\label{collisionssec}
In order to calculate the viscous relaxation rate, we need  the
particle-particle scattering cross section. The differential cross
section $d\sigma/d\Omega$ for $s$ wave scattering is related to
the scattering matrix ${\cal{T}}$ by
\begin{equation}
\frac{d\sigma}{d\Omega}=\frac{m^2|{\cal{T}}|^2}{(4\pi\hbar^2)^2}.\label{crosssect}
\end{equation}
An often used approximation close to a resonance is to use the
so-called unitarized vacuum scattering matrix
\begin{equation}
{\cal{T}}_{\rm uni}=\frac{4\pi\hbar^2 a}{m}\frac{1}{1+iqa}
\label{Tunitary}
\end{equation}
where $a$ is the  scattering length and $\hbar q$  the relative
momentum of the scattering particles. In the present paper, we
shall extend the calculation carried out in I of the viscous
relaxation rate to cover the temperature region near $T_c$ where
medium effects absent in the approximation (\ref{Tunitary}) turn
out to be important.

Close to a Feshbach resonance, the scattering length can be
written in the phenomenological form
\begin{equation}
a=a_{\rm bg}\left(1-\frac{\Delta B}{B-B_0}\right). \label{ares}
\end{equation}
Here, $\Delta B$ is the width of the resonance, $a_{\rm bg}$ the
background scattering length coming from non-resonant scattering
processes, and $B_0$ the position of the resonance. For atomic
gases close to a Feshbach resonance, a multi-channel effective
theory for atom-atom scattering  was recently
presented~\cite{GMBEvgeni}. The scattering matrix including medium
effects was shown to be given by
\begin{equation}
{\cal{T}}(\omega,{\mathbf{K}})=\frac{{\cal{T}}_{\rm
bg}}{\left(1+\frac{\Delta\mu\Delta B} {\hbar
\tilde{\omega}+h(\omega,{\mathbf{K}})-\Delta\mu(B-B_0)}\right)^{-1}
-{\cal{T}}_{\rm bg}\Pi(\omega,{\mathbf{K}})}, \label{T(mu,abg)}
\end{equation}
where $\hbar {\mathbf{K}}$ is the center-of-mass momentum of the
scattering particles and $\omega$ the frequency. Here
$\tilde{\omega}=\omega-\hbar K^2/4m$ and ${\cal{T}}_{\rm bg}=4\pi
a_{\rm bg}\hbar^2/m$, while $\Delta\mu$ is the magnetic moment of
the Feshbach molecule. The pair propagator $\Pi$ is
\begin{gather}
\Pi(\omega,{\mathbf{K}})=\int
\frac{d^3q'}{(2\pi)^3}
\left[\frac{1-f^0(\frac12\mathbf{K}-\mathbf{q'})-f^0(\frac12\mathbf{K}+\mathbf{q'})}
{\hbar \omega+i\delta-\frac{\hbar^2K^2}{4\,m}-\frac{\hbar^2q'^2}{m}}\right.\nonumber\\
\left.+\frac{1}{\hbar^2 q'^2/m-i\delta}\right] \label{Piopen}
\end{gather}
where $f^0$ is the equilibrium Fermi function. 
The quantity $h(\omega,{\mathbf{K}})$ in (\ref{T(mu,abg)})
describes effects coming from the fact that the Feshbach state is
a composite two-fermion object, but it can be neglected for most
resonances of experimental interest~\cite{GMBEvgeni} and we
therefore set it equal to zero, $h(\omega,{\mathbf{K}})=0$. For
the calculation of transport coefficients such as the viscosity,
the scattering matrix has to be evaluated on-shell with
$\hbar\omega$ equal to the energy of the two scattering particles,
i.e.\ $\hbar \omega=\hbar^2 K^2/4m+\hbar^2 q^2/m$. In a vacuum,
the on-shell version of (\ref{T(mu,abg)}) with 
$h(\omega,{\mathbf{K}})=0$ reduces to
\begin{gather}
{\cal{T}}= \frac{\frac{4\,\pi\hbar^2}{m}\,a_{\rm bg}}{\left(1
+\frac{\Delta\mu\Delta B}{\hbar^2
q^2/m-\Delta\mu(B-B_0)}\right)^{-1}+ia_{\rm bg}q}. \label{Tvacuum}
\end{gather}
We shall  use (\ref{T(mu,abg)}) for  calculating the
low-temperature viscosity for a resonantly interacting gas.

The approximation  (\ref{Tunitary}) for the scattering matrix
neglects two effects as compared to the many-body expression
(\ref{T(mu,abg)}). Firstly, even the vacuum limit (\ref{Tvacuum})
of (\ref{T(mu,abg)}) is not identical to (\ref{Tunitary}); they
agree only when the $\hbar^2q^2/m$ term in (\ref{Tvacuum}) can be
neglected. This term corresponds to an effective range of the
atom-atom interaction. Here we consider only broad resonances such
as the one at $B\simeq 201.6$G for $^{40}$K or the one at $B
\simeq 830$G for $^6$Li, for which the effective range can be
safely neglected~\cite{GMBEvgeni,GMBUni}.

Secondly, the approximation given by (\ref{Tunitary})  neglects
medium effects, which are included in (\ref{T(mu,abg)}) via the
propagator (\ref{Piopen}). These effects come from the occupation
of open-channel states. One effect of the medium is to shift the
resonance position away from its vacuum value
$B_0$~\cite{GMBEvgeni}. More importantly for the present purpose,
medium effects cause a significant increase in the scattering rate
for $T\rightarrow 0$~\cite{GMBPethick} due to the Fermi blocking
of the pair states into which the molecular state can decay. The
decay of the molecular state in a vacuum is the origin of  the
$iqa$ terms in the denominator of (\ref{Tunitary}) and
(\ref{Tvacuum}). These terms are replaced by ${\cal{T}}_{\rm
bg}\Pi$ in (\ref{T(mu,abg)}) when  medium effects are taken into
account. At low temperatures, this term is significantly reduced
from the vacuum value due to the Fermi-blocking factors in
(\ref{Piopen}); the lifetime of the resonant state is increased
leading to a corresponding increase in the scattering rate. We
shall demonstrate that this effect is significant close to the
superfluid transition temperature $T_c$.

\section{Homogeneous system}\label{HomSec}
We now present results for the viscosity of a homogeneous system.
The viscosity is calculated as a function of temperature for a
constant density $n$. We calculate the thermodynamic potential
$\Omega$ for fixed $n$ within the ladder approximation as a
function of temperature, taking into account two-body scattering
processes with a scattering matrix given by (\ref{T(mu,abg)}). The
chemical potential $\mu$ is determined by the condition
$-\partial_\mu\Omega=n$.
Our calculation of $\Omega$ is identical to the one described in
Ref.\ \cite{GMBUni} where more details are given. Once the
chemical potential is determined, we calculate $\eta$ using Eq.\
(16) of I with the cross section given  by (\ref{crosssect}) and
(\ref{T(mu,abg)}). The calculation of the chemical potential and
the viscosity requires multi-dimensional integrations which are
performed numerically using two different methods to check the
accuracy. The first method uses convoluted 1D integration routines
and the other is based on multidimensional Monte-Carlo
integration. Both yield the same result for the chemical potential
and the viscosity within less than $1$\%, thus confirming the
accuracy of the numerical calculations.

\subsection{Low and high temperature limits}
Before we present any results, let us first discuss what can be
calculated analytically. For low temperatures $T/T_{\rm F}\ll 1$
($kT_{\rm F}=\hbar^2k_{\rm F}^2/2m$), the Fermi factors in the
collision integral (see Eq.\ (13) in I) restrict the available
phase for the scattering particles such that their momenta are
close to the Fermi surface. This simplifies the integrations and
assuming the simple unitarized vacuum form (\ref{Tunitary}) for
the scattering matrix (${\cal{T}}={\cal{T}}_{\rm uni}$), an
analytical result for the viscosity can be obtained using standard
Fermi liquid tricks~\cite{ChrisFermiLiquidBook}. A lengthy but
straightforward analysis yields for $k_{\rm F}|a|\ll 1$ and
$T\rightarrow 0$
\begin{equation}
\eta=\frac{3n\hbar}{8\pi(k_{\rm F}a)^2}\frac{\epsilon_{\rm
F}^2}{(kT)^2} \label{ViscLowT}
\end{equation}
with $\epsilon_{\rm F}=\hbar^2k_{\rm F}^2/2m$. Equation
(\ref{ViscLowT}) is based on $T\ll T_{\rm F}$ and the assumption
${\cal{T}}= {\cal{T}}_{\rm uni}$. This assumption works well for
weak coupling with $T_c\ll T_{\rm F}$: there is then a temperature
regime for which $T\ll T_{\rm F}$ such that all scattering momenta
are confined to the Fermi surface, and $T\gg T_c$ such that medium
effects can be neglected. However, for strong coupling $T_c$
becomes comparable to $T_{\rm F}$ and there is no temperature
regime for which $T\ll T_{\rm F}$ and  $T\gg T_c$.

In the classical limit $T\gg T_{\rm F}$, an analytical expression
for the viscosity can also be obtained using ${\cal{T}}=
{\cal{T}}_{\rm uni}$:
\begin{gather}
\eta=\frac{5\sqrt{\pi mkT}}{8\bar{\sigma}}\nonumber\\
=\frac{5\sqrt{mkT}}{32\sqrt{\pi}}\times\left\{\begin{array}{lcl}
a^{-2}   &,&T\ll T_a\\
3mkT/\hbar^2  &,&T\gg T_a
\end{array}
\right.
\label{ViscHighT}
\end{gather}
with $kT_a=\hbar^2/(ma^2)$ and
 \begin{equation}
\bar{\sigma}=\frac{4\pi a^2}{3}\int_0^{\infty}dx
x^7e^{-x^2}(1+x^2T/T_a)^{-1} \label{barcross}
\end{equation}
the momentum averaged cross section. As opposed to the
low-temperature limit, the assumption ${\cal{T}}= {\cal{T}}_{\rm
uni}$ is reasonable for $T\gg T_{\rm F}$, even in the
strong-coupling case, since medium effect on the scattering matrix
are suppressed at high temperatures.

\subsection{Numerical results}
\begin{figure}
\includegraphics[width=\columnwidth,angle=0,clip=]{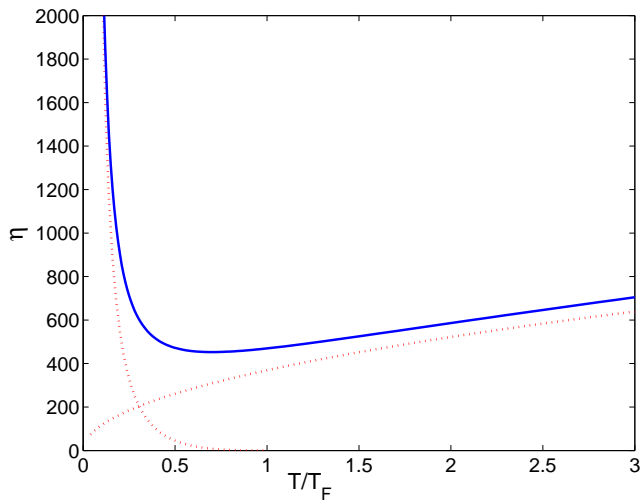}
\caption{(Color online) The viscosity $\eta$ in units of $n\hbar$
for background scattering. The solid line is the numerical result
with the scattering matrix (\ref{T(mu,abg)}), the dotted lines are
the low  and high temperature limits.} \label{BackgroundVisc}
\end{figure}

 Fig.\ \ref{BackgroundVisc} shows the
viscosity $\eta$ in the weak-coupling limit $k_{\rm F}|a|\ll 1$.
We have chosen parameters in (\ref{T(mu,abg)}) corresponding to a
generic weak interaction characterized by background scattering
with $k_{\rm F} a_{\rm bg}=-0.1$, $\Delta B/(B-B_0)=0.0012$, and a negligibly small effective
range.
For weak interaction, the scattering matrix reduces to
$4\pi\hbar^2 a_{\rm bg}/m$ and there are no effects of the medium
as long as $T$ is well above  $T_c\ll T_{\rm F}$. In fact, the
solid line in Fig.\ \ref{BackgroundVisc} is indistinguishable from
the result obtained if we had used $4\pi\hbar^2 a_{\rm bg}/m$ for
the scattering matrix of (\ref{T(mu,abg)}) as expected. We see
that the viscosity approaches the low- and high-temperature forms
given by (\ref{ViscLowT}) and (\ref{ViscHighT}) as expected, with
a minimum located at $T\approx 0.7 T_{\rm F}$.

\begin{figure}
\includegraphics[width=\columnwidth,angle=0,clip=]{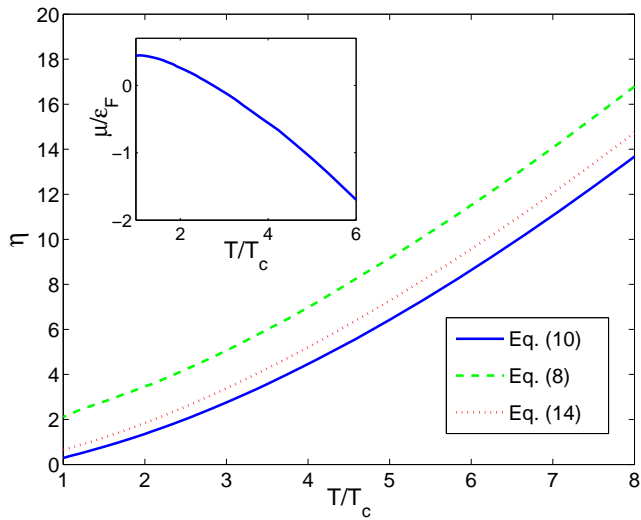}
\caption{(Color online) The viscosity $\eta$ in units of $n\hbar$
for resonant scattering as a function of temperature for $T\ge
T_c$. The solid line is the numerical result with the full
scattering matrix (\ref{T(mu,abg)}), the dashed line the numerical
result using the vacuum approximation (\ref{Tunitary}). The dotted
line is the low-temperature result (\ref{ViscHighT}). The inset
shows the chemical potential.} \label{UniVisc}
\end{figure}

Fig.\ \ref{UniVisc} shows the viscosity $\eta$ in the
strong-coupling limit.
We have chosen parameters in (\ref{T(mu,abg)}) corresponding to a
generic resonant interaction characterized by $k_{\rm F}a=-11.8$
and a negligible effective range.
 For this set of parameters very close to resonance, we find
the critical temperature for the superfluid transition to be
$T_c\approx 0.26T_{\rm F}$ in excellent agreement with earlier
numerical BEC-BCS crossover results based on a similar
approximation for the thermodynamic potential~\cite{BECBCS,Pieri}.

Just above the critical temperature $T_c$, the viscosity is
reduced almost by an order of magnitude (a factor $\approx 7.5$)
as compared to the value obtained using the vacuum scattering
matrix. The reduction reflects the increase in the scattering rate
due to Fermi blocking of the molecule decay as explained in Sec.\
\ref{collisionssec}. These medium effects cause the zero
center-of-mass (COM) momentum ${\cal{T}}$ matrix to diverge at
$T_c$, which is the signature of the Cooper instability. Thus, we
expect the increase in the scattering rate due to the medium to be
significant for temperatures close to $T_c$, regardless of the
value of the coupling strength. This is confirmed by numerical
calculations for different values of the coupling: Medium effects
on the scattering rate and thus the viscosity are significant when
\begin{equation}
\frac{T-T_c}{T_c}\lesssim 1.
\end{equation}
In the high temperature  limit on the other hand, the medium
effects on the scattering are suppressed as expected and the
viscosity approaches its classical value as can be seen from Fig.\
\ref{UniVisc}.
We conclude that the medium increases the scattering rate
significantly as compared to the vacuum value for temperatures
close to $T_c$ due to the Cooper instability.

Note that since $kT_c/\mu(T_c)\approx 0.6$  is rather large for
strong coupling, the $T^{-2}$ increase in the viscosity coming
from the phase space blocking in the collision integral is not
observed for $T\ge T_c$ in Fig.\ \ref{UniVisc}.


The above results show that the minimum viscosity for a Fermi gas
in the normal phase for $k_{\rm F}|a|\rightarrow\infty$ and
$T=T_c$ is
\begin{equation}
\eta_{\rm min}=\alpha n\hbar \label{Lowerbound}
\end{equation}
with $\alpha\approx 0.15$. This minimum is significantly less than
the value $\alpha\approx 1.1$ obtained when the vacuum scattering
matrix given by (\ref{Tunitary}) is used. We caution that the
minimum value of $\eta/n\hbar$ obtained in (\ref{Lowerbound})
depends on the validity of our starting point, the semiclassical
kinetic equation. A simple relaxation-time approximation to the
collision integral, yielding the low-temperature viscosity
$\eta\approx mnv_{\rm F}^2\tau/5$, would result in $\eta\approx
0.4 n\hbar$ if $\tau=\hbar/\epsilon_{\rm F}$. However, for such
small values of $\tau$ the semiclassical kinetic equation may need
modification due to the spectral broadening of the momentum
states. It is interesting to note that a related problem, the
viscosity of a hot quark-gluon plasma, was recently analyzed
within the Kubo formalism \cite{Peshier} which yielded results
that differed significantly from those obtained from the Boltzmann
equation.


\subsection{Effects of superfluidity}

The viscosity of a superfluid gas of fermions has been
investigated extensively in the context of superfluid
$^3$He~\cite{VW}. Below the transition temperature $T_c$,
 the viscosity drops rapidly as a function of
temperature, the relative decrease in viscosity being of order
$\Delta/kT_c$. The physical reason for the large drop is the
change in the quasiparticle dispersion relation and the
modification of the collision operator. For the purpose of
illustration we shall here discuss the two effects separately,
although they must of course  be treated together in a consistent
calculation.

Firstly, the group velocity is changed in the superfluid state,
since the dispersion relation of Bogoliubov quasiparticles is
\begin{equation}
E=\sqrt{\xi^2+\Delta^2},
\end{equation}
where $\xi=p^2/2m-\mu$ and $\Delta$ is the energy gap. The
group velocity ${\bf v}_g=\partial E/\partial {\bf p}$ in the
superfluid state is thus
\begin{equation}
{\bf v}={\bf \hat{p}}v_{\rm F}\frac{\xi}{E},
\end{equation}
where  ${\bf \hat{p}}$ is a unit vector along ${\bf p}$ and
$v_{\rm F}=\hbar k_{\rm F}/m$ is the Fermi velocity. This affects
the velocity component $v_y$ appearing on the left hand side of
the kinetic equation and in the expression  for the momentum
current density. To illustrate the effect of the change in group
velocity, let us make  a relaxation time approximation to the
collision operator occurring in the Boltzmann equation. The shear
viscosity is obtained by assuming a flow velocity ${\bf u}$ of the
form $ {\bf u}=(u_x(y),0,0)$ and linearizing in the gradient
$\partial u_x/\partial y$, as a result of which the kinetic
equation for the non-equilibrium distribution function $f$ becomes
\begin{equation} -\frac{\partial
u_x}{\partial y}v_yp_x\frac{\partial f^0}{\partial
\epsilon}=-\frac{f-f^0}{\tau}. \label{linB1}
\end{equation}
The shear viscosity $\eta$ relates the momentum current density
$\Pi_{xy}$, given by
\begin{equation}
\Pi_{xy}=2\int\frac{d{\bf p}}{(2\pi\hbar)^3}v_yp_xf
,\label{momcur}
\end{equation}
to the gradient of the flow velocity according to
$\Pi_{xy}=-\eta\partial u_x/\partial y$. Inserting the resulting
distribution function in (\ref{momcur}) and carrying out the
integration over angles we can perform the energy integration
analytically, using the fact that
\begin{equation}
\frac{\xi^2}{E^2}=1-\frac{\Delta^2}{\xi^2+\Delta^2}\simeq
1-\pi\Delta\delta(\xi).\label{deltaid}
\end{equation}
Here the representation of the Lorentzian by a delta-function is
valid as long as the remaining part of the integrand varies slowly
on the scale of $\Delta$, that is provided $\Delta\ll kT$. Since
the value of $f^0(1-f^0)$ at the chemical potential is $1/4$ for a
degenerate fermion gas we obtain
\begin{equation}
\eta=\eta_n\left(1-\frac{\pi}{4}\frac{\Delta}{kT_c}\right),\label{reltimeapp}
\end{equation}
to lowest order in $\Delta/kT_c$. The quantity $\eta_n=mnv_{\rm
F}^2\tau/5$ is the viscosity in the normal state at $T_c$.

Secondly, the probability of two-quasiparticle scattering
processes is modified by the coherence factors appearing in the
Bogoliubov transformation. Since the number of quasiparticles is
not a conserved quantity in the superfluid state, it is also
necessary to include processes in which one quasiparticle decays
into three as well as the inverse processes.

Taking  both these effects  into account, it is possible near the
transition temperature to express the relative decrease in
viscosity in terms of normal-state quantities and relate the
coefficient in front of $\Delta/kT_c$ to the ratio between the
viscous relaxation time $\tau_{\eta}$ and the lifetime $\tau(0)$
of a quasiparticle in the normal state at the Fermi energy. One
finds~\cite{Bhatta} that the change in viscosity is given by
\begin{equation}
\frac{\delta \eta}{\eta}=
-\frac{\pi}{4}\left(1-\frac{\pi^2}{12}+
\frac{\tau_{\eta}}{\tau(0)}\right)^2\frac{\tau(0)}{\tau_{\eta}}\frac{\Delta}{kT_c}.
\end{equation}
The coefficient of the $\Delta/kT_c$ term may thus be expressed
solely in terms of normal-state quantities. For the trial function
(\ref{X}) used here, the ratio between the lifetime and the
viscous relaxation time is~\cite{HenrikBook}
\begin{equation}
\frac{\tau(0)}{\tau_{\eta}}=2\left({\int_0^1dx\frac{x^5}{\sqrt{1-x^2}}\frac{d\sigma}{d\Omega}}\right)
\left({\int_0^1dx\frac{x}{\sqrt{1-x^2}}\frac{d\sigma}{d\Omega}}\right)^{-1}.\label{wear}
\end{equation}
The medium cross section diverges at zero COM momentum at $T_c$. This leads to
a non-integrable pole at $x=1$ in both integrals in (\ref{wear}) and therefore
\begin{equation}
\frac{\tau(0)}{\tau_{\eta}}=2.\label{wear1}
\end{equation}
This is a universal result, independent of the coupling strength,
coming from the Cooper instability at $T_c$.

The viscous relaxation time is obtained from the viscosity by
removing the factor $(X,X)$, where $X$ is the trial function
proportional to $v_yp_x$. By the same reasoning that led to
(\ref{reltimeapp}) it follows that the value $(X,X)_s$ of the
scalar product in the superfluid state is related to its value
$(X,X)_n$ in the normal state by
 \begin{equation}
(X,X)_s\simeq
(X,X)_n\left(1-\frac{\pi}{4}\frac{\Delta}{kT_c}\right)
 \end{equation}
to leading order in $\Delta/kT_c$. Consequently, the change in
$\tau_{\eta}$ is  given by
\begin{equation}
\frac{\delta
\tau_{\eta}}{\tau_{\eta}}=0.0643\frac{\Delta}{kT_c}.\label{final}
\end{equation}
Due to the smallness of the numerical factor in (\ref{final}) we
conclude that there is very little  change in the viscous
relaxation rate upon entering the superfluid phase.

\section{Trapped Gas}
We now present results for the thermal relaxation rate as a
function of temperature both in the weak and the resonant coupling
limit for a trapped two-component Fermi gas.
To keep the number of trapped particles $N=2N_\sigma$ constant
(there are $N_\sigma$ atoms in each hyperfine state), we calculate
$\Omega$ in the ladder approx.\ as in Sec.\ \ref{HomSec} and the
chemical potential is determined by the condition
$N=-\partial_\mu\Omega$. The relaxation rate is then calculated
from (\ref{ThermalRate}). The presence of the trapping potential
$V({\mathbf{r}})$ is treated in the Thomas-Fermi approximation by
a local chemical potential
$\mu({\mathbf{r}})=\mu-V({\mathbf{r}})$. This requires an
additional integration over the spatial coordinates
${\mathbf{r}}$, and the resulting multidimensional integrals are
evaluated using a Monte-Carlo routine.

\subsection{Low and high temperature limits}
As for the homogeneous case, the integrals in (\ref{ThermalRate})
for the thermal relaxation rate can be calculated analytically in
the low and high temperature limits using the unitarized vacuum
form ${\cal{T}}_{\rm uni}$ for the ${\cal{T}}$ matrix. For
$T/T_{\rm F}\ll 1$ and weak coupling $k_{\rm F}|a|\ll 1$, we
obtain
\begin{equation}
\frac{1}{\tau_T}=
\frac{16\pi (kT)^2}{15\hbar^3}ma^2.
\label{TrateLowT}
\end{equation}
For high temperatures, $T\gg T_{\rm F}$, we obtain
\begin{gather}
\frac{1}{\tau_T}=\frac{N}{10\pi^2}\frac{m\tilde{\omega}^3\bar{\sigma}}{kT}\nonumber\\
=\frac{2N\tilde{\omega}^3}{5\pi kT}\times
\left\{\begin{array}{lcl}
a^2  &,&T\ll T_a\\
\hbar^2/(3mkT) &,&T\gg T_a
\end{array}
\right.
\label{TrateHighT}
\end{gather}
with $\tilde{\omega}^3=\omega_x\omega_y\omega_z$ and
$\bar{\sigma}$ given by (\ref{barcross}). As for the homogeneous
case,  ${\cal{T}}={\cal{T}}_{\rm uni}$ is a good approximation at
low temperatures for weak coupling only,  whereas it is  valid at
high temperatures for all coupling strengths.

\subsection{Numerical results}
For weak coupling $k_{\rm F}|a|\ll 1$, the numerical calculations
confirm as expected that there  are no effects of the medium
(except for very small $T$ close to $T_c$) and the results for the
thermal relaxation rate are identical to the ones reported in
Fig.\ 2 of I.

\begin{figure}
\includegraphics[width=\columnwidth,angle=0,clip=]{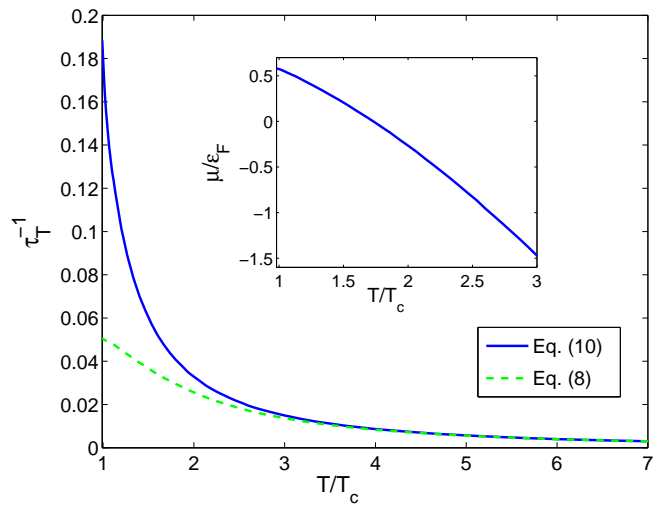}
\caption{(Color online) The thermal relaxation rate in a trap in
units of $\epsilon_{\rm F}/\hbar$ for the resonant coupling case
$k_{\rm F}|a|\gg 1$. The solid line is the numerical result with
the full scattering matrix in (\ref{T(mu,abg)}), the dashed line
the numerical result using the vacuum approximation
(\ref{Tunitary}) for the scattering matrix. The curves are given
for $T\ge T_c\approx 0.3T_{\rm F}$.} \label{TRate11}
\end{figure}

We therefore concentrate on the strongly interacting limit. Fig.\
\ref{TRate11} shows the thermal relaxation rate $\tau_T^{-1}$ as a
function of temperature for a strongly interacting system with
$k_{\rm F}|a|\gg 1$. The rate is in units of $\epsilon_{\rm
F}/\hbar$ where $\epsilon_{\rm F}=\hbar\tilde{\omega}(3N)^{1/3}$
is the Fermi energy for $N$ non-interacting trapped particles.
We have chosen parameters in (\ref{T(mu,abg)}) corresponding to a
resonant interaction with $k_{\rm F}a=-11.8$ and a small effective
range $k_{\rm F}r_{\rm eff}\ll 1$. As for the homogeneous case,
the  strong interaction gives rise to a superfluid transition
below a critical temperature $T_c$. We find $T_c\approx 0.3T_{\rm
F}$ at resonance ($k_{\rm F}|a|\gg 1$) in excellent agreement with
the results in Ref.\ \cite{Perali}.

We see from Fig.\ \ref{TRate11} by comparing the solid and the
dashed lines, that the medium increases the relaxation rate
significantly for low temperatures as compared to the vacuum
prediction: close to $T_c$ the rate is approximately $3.6$ times
higher than the vacuum result. The increase is smaller than for
the homogeneous case since the effects of the medium are reduced
near the edge of the cloud where the density is reduced. The
reason for the significant increase of the rate  is of course
identical to the effect discussed for the homogeneous case: the
medium increase of  the resonant scattering rate for low
temperatures $(T-T_c)/T_c\lesssim 1$ signals the Cooper
instability at $T_c$. For high temperatures, the medium effects
are negligible. As for the homogeneous case, the $T^2$ dependence
of the rate coming from the Fermi blocking of the available
scattering states for $T/T_{\rm F}\ll 1$ is not observed for $T\ge
T_c$ since $kT_c/\mu(T_c)\approx 0.5$ is rather high.

Fig.\ \ref{TRate11} shows the maximum relaxation rate for a
two-component trapped Fermi gas for $k_{\rm
F}|a|\rightarrow\infty$. The large difference between the
relaxation rate for a weakly interacting system and a resonantly
interacting system should be contrasted with the much smaller
effects of the resonant interaction on thermodynamic properties.

\subsection{Experiments}
Let us comment on the possible experimental observation of the
effects described above. The thermal relaxation rate can be
measured by ``heating'' the gas preferentially in one spatial
direction followed by observing the time evolution towards
equilibrium with a uniform temperature in all
directions~\cite{Loftus,Regal}. These experiments have already
demonstrated the saturation of the scattering rate at resonance in
the classical regime. By performing such experiments for
decreasing temperatures, one should observe a significant increase
in the rethermalization rate as compared to the vacuum rate when
$(T-T_c)/T_c\lesssim 1$.

\section{Conclusion}
Using a variational approach, we have calculated the viscosity and
the thermal relaxation rate of an interacting Fermi gas for weak
and for strong coupling. Both homogeneous and trapped systems
were considered and medium effects on the scattering
properties were taken into account. The rates were shown to be
significantly increased for temperatures close to the superfluid
transition temperature. This is due to the presence of the medium and
signals the onset of Cooper pairing at $T_c$. We furthermore
considered the effects of superfluidity below $T_c$ to leading
order in the energy gap. The  effects described in this paper can
be measured directly by rethermalization experiments.

\end{document}